\documentclass[12pt]{article}
\usepackage{graphicx}
\def \app{D_{\pi \pi}}
\def \b{{\cal B}}

\def \bea{\begin{eqnarray}}
\def \beq{\begin{equation}}

\def \cn{Collaboration}

\def \eea{\end{eqnarray}}
\def \eeq{\end{equation}}
\def \ite{{\it et al.}}

\def \s{\sqrt{2}}

\def \3half{\frac{3}{2}}

\begin{document}

\begin{flushright}
SLAC-PUB-11380 \\
hep-ph/0508047 \\
August 2005 \\
\end{flushright}

\renewcommand{\thesection}{\Roman{section}}
\renewcommand{\thetable}{\Roman{table}}
\centerline{\bf A precise sum rule among four $B\to K\pi$ CP asymmetries
\footnote{To be published in Physics Letters B.}}
\bigskip
\centerline{Michael Gronau}
\medskip
\centerline{\it Stanford Linear Accelerator Center}
\centerline{\it Stanford University, Stanford, CA 94309}
\medskip
\centerline{and}
\medskip
\centerline{\it  Physics Department, Technion} 
\centerline{\it Haifa 32000, Israel}
\bigskip

\begin{quote}
A sum rule relation is proposed for direct CP 
asymmetries in $B\to K\pi$ decays. Leading 
terms are identical in the isospin symmetry limit, while 
subleading terms are equal in the flavor SU(3) and heavy quark 
limits. The sum rule predicts $A_{\rm CP}(B^0\to K^0\pi^0) = -0.17 
\pm 0.06$ using current asymmetry measurements for the other 
three $B\to K\pi$ decays. A violation of the sum rule would be 
evidence for New Physics in $b\to s\bar q q$ transitions.  
\end{quote}

\leftline{\qquad PACS codes:  12.15.Hh, 12.15.Ji, 13.25.Hw, 14.40.Nd}

\bigskip

CP asymmetry measurements in neutral $B$ decays involving 
an interference between $B^0$--$\bar B^0$ mixing and $b\to c\bar c s$ or 
$b\to u\bar ud$ transitions improve our knowledge of the 
Cabibbo-Kobayashi-Maskawa (CKM) phases $\beta\equiv 
{\rm Arg}(-V^*_{cb}V_{cd}/V^*_{tb}V_{td})$ and $\alpha\equiv 
{\rm Arg}(-V^*_{tb}V_{td}/V^*_{ub}V_{ud})$ beyond information obtained 
from all other CKM constraints~\cite{MGSak}. While time-dependent 
asymmetries in $b\to s\bar q q$ transitions ($q=u,d,s$) indicate
a potential deviation from $\sin 2\beta$~\cite{beta}, the current 
statistical significance of the discrepancy is insufficient for 
claiming a serious anomaly.

Extraction of the  weak phase $\gamma\equiv {\rm Arg}
(-V^*_{ub}V_{ud}/V^*_{cb}V_{cd})$ from the direct CP asymmetry 
measured recently in $B^0\to K^+\pi^-$~\cite{K+pi-Ba,K+pi-Be} is obstructed 
by large theoretical uncertainties in strong interaction phases. 
Direct CP asymmetries can provide evidence for New Physics in $B^+\to 
\pi^+\pi^0$, where the Standard Model predicts a vanishing asymmetry, 
including tiny electroweak penguin contributions~\cite{GPY}. 
Other tests based on direct asymmetries, which require studying carefully 
U-spin symmetry breaking effects, are provided by pairs of processes, e.g. 
$B^+\to K^0\pi^+$ and $B^+\to \bar{K^0}K^+$ or $B^0\to K^+\pi^-$ and 
$B_s\to \pi^+K^-$, in which CP asymmetries are related by U-spin 
symmetry interchanging $d$ and $s$ quarks~\cite{Uspin}. Precision 
tests would be provided 
by CP  asymmetry relations, in which isospin relates dominant terms in the 
asymmetries while flavor SU(3) relates subdominant terms. 
The motivation for this study is proposing such a relation among 
$B\to K\pi$ asymmetries, a violation of which could serve as an alternative 
clue for physics beyond the Standard Model in $b\to s \bar q q$ transitions. 

Direct CP asymmetries in all four $B\to K\pi$ decay processes, measured by 
the Babar~\cite{Babar} and Belle~\cite{Belle} collaborations, are quoted 
in Table I together with their averages. (We do not quote earlier CLEO 
measurements which involve considerably larger errors.) One defines 
by convention
\beq
A_{CP}(B \to f) \equiv \frac{\Gamma(\bar B \to \bar f) - \Gamma(B\to f)}
{\Gamma(\bar B \to \bar f) + \Gamma(B \to f)}~.
\eeq
A nonzero asymmetry was measured in $B^0\to K^+\pi^-$, $A_{\rm CP}=
(-12.0 \pm 1.9)\%$, where the experimental error is smallest among the 
four $B\to K\pi$ decays. The other 
three asymmetries, of which that in $B\to K^0\pi^0$ involves the 
largest experimental error, are consistent with zero.
Table I quotes also for later use corresponding CP-averaged branching ratios 
in units of $10^{-6}$~\cite{HFAG}, including a very recent Babar measurement 
of $\b(B^0\to K^+\pi^-)$~\cite{Ford}. 

\begin{table}
\caption{CP asymmetries $A_{CP}$ for $B \to K \pi$ decays. In parentheses 
are corresponding branching ratios in units of $10^{-6}$.
\label{tab:Kpiasym}}
\begin{center}
\begin{tabular}{c c c c} \hline \hline
Decay mode & Babar~\cite{Babar} & Belle~\cite{Belle} & Average \\ \hline
$B^0\to K^+ \pi^-$ & $-0.133\pm 0.030 \pm 0.009$ & $-0.113\pm 
0.022\pm 0.008$ & $-0.120 \pm 0.019$ \\
  & $(19.2 \pm 0.6 \pm 0.6)$ & $(18.5 \pm 1.0 \pm 0.7)$ & ($18.9 \pm 0.7$)\\
$B^+\to K^+ \pi^0$ & $0.06\pm 0.06\pm 0.01$ &  $0.04 \pm 0.04\pm 0.02$ 
& $0.05 \pm 0.04$ \\
 & $(12.0 \pm 0.7 \pm 0.6)$ & $(12.0 \pm 1.3^{+1.3}_{-0.9})$ & 
$(12.1 \pm 0.8)$ \\
$B^0\to K^0 \pi^0$ & $-0.06\pm 0.18\pm 0.03$ & $0.11 \pm 0.18\pm 0.08$ 
& $0.02 \pm 0.13$ \\
 & $(11.4 \pm 0.9 \pm 0.6)$ & $(11.7 \pm 2.3^{+1.2}_{-1.3})$ & 
$(11.5 \pm 1.0)$ \\
$B^+\to K^0 \pi^+$ & $-0.09\pm 0.05 \pm 0.01$ & $0.05\pm 0.05\pm 0.01$ & 
$-0.02\pm0.04$ \\
 & $(26.0  \pm 1.3 \pm 1.0)$ & $(22.0 \pm 1.9 \pm 1.1)$ & $(24.1 \pm 1.3)$ \\ 
\hline \hline
\end{tabular}
\end{center}
\end{table}

The purpose of this Letter is to prove a sum rule
among the four $B\to K\pi$ CP rate differences,
\beq\label{DSR}
\Delta(K^+\pi^-) + \Delta(K^0\pi^+) \approx 2\Delta(K^+\pi^0) + 
2\Delta(K^0\pi^0)~,
\eeq
where we define
\beq\label{Deltadef}
\Delta(B\to f) \equiv \Gamma(\bar B\to \bar f) - \Gamma(B\to f)~.
\eeq
This sum rule, reminiscent of a similar 
sum rule among partial decay rates~\cite{GR99,Lipkin}, is expected 
to hold within an accuracy of several percent. 
Using the approximation (see branching ratios in Table I and the 
discussion in the paragraph below Eq.~(\ref{C})),
\beq\label{Gamma}
\Gamma(K^+\pi^-) \approx \Gamma(K^0\pi^+) \approx 
2\Gamma(K^+\pi^0) \approx 2\Gamma(K^0\pi^0)~,
\eeq
this implies at a somewhat lower precision a sum rule among CP 
asymmetries,
\beq\label{SR}
A_{\rm CP}(K^+\pi^-) + A_{\rm CP}(K^0\pi^+) \approx
A_{\rm CP}(K^+\pi^0) + A_{\rm CP}(K^0\pi^0)~.
\eeq 
The equality of leading terms in the sum rule (\ref{DSR}) will be 
shown to follow from isospin symmetry, while subleading terms are 
equal in the flavor SU(3) and heavy quark limits. For the most part, we 
will make no assumption relating $B\to K\pi$ decays to $B\to\pi\pi$ 
decays. 

A somewhat less precise relation excluding $A_{\rm CP}(K^0\pi^+)$ in 
(\ref{SR}), which holds under more restricted conditions, was proposed 
recently in a broader context~\cite{bspeng}. An oversimplified but too 
crude relation,
\beq\label{naive}
A_{\rm CP}(K^+\pi^-) \sim A_{\rm CP}(K^+\pi^0)~,
\eeq
was suggested several years ago~\cite{GR99} making too strong an
assumption about color suppressed tree amplitudes.
The latter relation, which is quite far from what is being measured
(see Table I), 
has recently provoked discussions about an anomalously large 
color-suppressed amplitude, an enhanced electroweak penguin amplitude
and possible New Physics effects~\cite{NP}. 

Let us recapitulate for completeness the structure of 
hadronic amplitudes in charmless $B$ decays, specifying carefully our 
assumptions. The effective Hamiltonian governing $B\to K\pi$ decays is 
given by~\cite{BBL}
\beq\label{H}
{\cal H}_{\rm eff} = -\frac{4G_F}{\s}\left [\sum_{U=u,c}\lambda_U(c_1O^U_1 + 
c_2O^U_2) - \lambda_t\sum_{i=3}^{10}c_iO_i \right ].
\eeq
where $\lambda_U = V^*_{Ub}V_{Us}, \lambda_t\equiv V^*_{tb}V_{ts}$.
The dozen operators $O^U_j$ and 
$O_i$ are four-quark operators, with given flavor and chiral structure, 
including current-current operator $O^U_{1,2}$, QCD-penguin operators $O_i, 
i=3-6$, and electroweak penguin (EWP) operators $O_i, i=7-10$. The real
Wilson coefficients, which were calculated beyond the leading logarithmic 
approximation, are $c_1 \approx 1.10,~c_2 \approx -0.20,~c_{3-6}\sim 
{\rm few}\times10^{-2},~c_{7,8}\sim {\rm few}\times10^{-4},
c_9 \approx -0.010$ and $c_{10} \approx 0.0020$. 
Contributions of $O_7$ and $O_8$ can be safely 
neglected, as one does not expect a huge enhancement of their hadronic 
matrix elements relative to those of $O_9$ and $O_{10}$. The latter 
operators, involving larger Wilson coefficients, have a (V-A)(V-A) 
structure similar to the current-current operators. 

All four quark operators can be written as a sum of SU(3) representations, 
${\overline {\bf 15}}$, ${\bf 6}$ and ${\bar {\bf 3}}$, into which the 
product ${\bar {\bf 3}}\otimes {\bf 3}\otimes {\bar {\bf 3}}$ can be 
decomposed~\cite{GHLR,GrLe}. Current-current and EWP operators 
which involve the same (V-A)(V-A) structure consist of identical 
SU(3) operators. Thus one finds simple proportionality relations between 
current-current (here denoted by a subscript $T$ for ``tree'') and EWP 
operators belonging to $\overline {\bf 15}$ and ${\bf 6}$ representations
\cite{GPY,NR,VP},
\bea\label{15}
{\cal H}_{EWP}(\overline{\bf 15}) &=& -\frac32 \frac{c_9+c_{10}}{c_1+c_2} 
\frac{\lambda_t}{\lambda_u} 
{\cal H}_T(\overline{\bf 15})~,\\
\label{6}
{\cal H}_{EWP}({\bf 6}) &=& \frac32 \frac{c_9-c_{10}}{c_1-c_2}
\frac{\lambda_t}{\lambda_u}{\cal H}_T({\bf 6})~.
\eea 
These operator relations have useful consequences in $B\to K\pi$ decay 
amplitudes. The first relation was applied in~\cite{NR} and both relations
were used in~\cite{GPY}. In the following discussion we will apply SU(3) 
to the subdominant EWP amplitudes by using Eqs.~(\ref{15})--(\ref{6}). 
Dominant terms in $B\to K\pi$ asymmetries will be shown to be related by 
isospin symmetry alone.

The effective Hamiltonian (\ref{H}) permits a general decomposition of 
the four $B\to K\pi$ amplitudes into terms of distinct topologies 
representing hadronic matrix elements of corresponding 
operators in (\ref{H}). Using the unitarity of the CKM matrix,
$\lambda_u +\lambda_c + \lambda_t =0$, and defining $P_{tc}\equiv 
P_t - P_c, P_{uc}\equiv P_u - P_c$, one has~\cite{GHLR}:     
\bea\label{+-}
-A(K^+\pi^-) & = & \lambda_u(P_{uc} + T) 
+ \lambda_t(P_{tc} + \frac{2}{3}P^C_{EW})~,\\
-\s A(K^+\pi^0) & = & \lambda_u(P_{uc} + T + C + A)  
+ \lambda_t(P_{tc} + P_{EW} + \frac{2}{3}P^C_{EW})~,\\
\s A(K^0\pi^0) & = & \lambda_u(P_{uc} - C) 
+ \lambda_t(P_{tc} - P_{EW} - \frac{1}{3}P^C_{EW})~,\\
\label{0+}
A(K^0\pi^+) & = & \lambda_u(P_{uc} + A)  
+ \lambda_t(P_{tc} - \frac{1}{3} P^C_{EW})~.
\eea
The amplitudes $P_u, T, C, A$ and $P_c$ are contributions from the first 
sum in (\ref{H}), corresponding to $U=u$ and $U=c$, respectively,
while $P_t, P_{EW}$ and $P_{EW}^c$ originate from the second sum.
The terms $P, T, C$ and $A$ represent penguin, color-allowed tree, 
color-suppressed tree and annihilation topologies, respectively.
Specific EWP contributions were expressed in terms of color-allowed 
and color suppressed amplitudes, $P_{EW}$ and $P^C_{EW}$, using a 
simple substitution~\cite{GHLR},
\beq
\lambda_u C \to \lambda_u C + \lambda_t P_{EW}~,~~~
\lambda_u T \to \lambda_u T + \lambda_t P^C_{EW}~,~~~
\lambda_u P_{uc} \to \lambda_u P_{uc} - \frac{1}{3} \lambda_t P^C_{EW}~.
\eeq

The four physical amplitudes can also be decomposed into three isospin 
amplitudes~\cite{isospin}, a contribution $B_{1/2}$ with $I(K\pi)=1/2$ 
from the isosinglet 
part of ${\cal H}_{\rm eff}$, and two amplitudes $A_{1/2,3/2}$ with $I(K\pi) 
=1/2,3/2$ from the isotriplet part of ${\cal H}_{\rm eff}$:
\bea\label{I-1}
-A(K^+\pi^-) & = & B_{1/2} - A_{1/2} - A_{3/2}~,\\
-\s A(K^+\pi^0) & = & B_{1/2} + A_{1/2} - 2A_{3/2}~,\\ 
\s A(K^0\pi^0) & = & B_{1/2} - A_{1/2} + 2A_{3/2}~,\\
\label{I-4}
A(K^0\pi^+) & = & B_{1/2} + A_{1/2} + A_{3/2}~.
\eea
One has
\bea\label{B}
B_{\frac{1}{2}} & = & \lambda_u\left [P_{uc} + \frac{1}{2}(T+A)\right]
+\lambda_t\left [P_{tc} + \frac{1}{6}P^C_{EW}\right ]~,\\
A_{\frac{1}{2}} & = & -\lambda_u \frac{1}{6}\left [T - 2C - 3A\right ] + 
\lambda_t \frac{1}{3}\left [ P_{EW} - \frac{1}{2}P^C_{EW}\right ]~,\\
\label{A3/2}
A_{\frac{3}{2}} & = & -\lambda_u \frac{1}{3}\left [ T+ C \right ] - 
\lambda_t \frac{1}{3}\left [ P_{EW} + P^C_{EW}\right ]~. 
\eea

Eqs.~(\ref{+-})-(\ref{0+}) are quite general, providing a common basis 
for QCD calculations of $B\to K\pi$ amplitudes~\cite{BBNS,KLS}. The terms 
in parentheses involve magnitudes of hadronic amplitudes and strong 
interaction phases, which are hard to calculate without making further 
assumptions. 
For instance, the term $P_c$ may involve sizable long 
distance ``charming penguin'' contributions which must be fitted to the 
data~\cite{Charmpeng}. Our following arguments will be independent of 
specific hadronic calculations, relying mainly on isospin and flavor SU(3) 
symmetry properties of certain terms. 
SU(3) breaking corrections will be estimated using generalized 
factorization~\cite{MN}.  

We will use
Eqs.~(\ref{15}) and (\ref{6}), which imply approximate SU(3) relations 
between $P_{EW}, P^C_{EW}$, on the one hand, and $T$ and $C$, on the 
other~\cite{GPY,NR},
\bea\label{T+C}
P_{EW} + P^C_{EW} & \approx & -\frac{3}{2}\frac{c_9 + c_{10}}{c_1 + c_2}
(T+C)~,\\
\label{C}
P^C_{EW} & \approx & - \frac{3}{2}\frac{c_9 + c_{10}}{c_1 + c_2}C~.
\eea
In the second equation we used $(c_9 - c_{10})/(c_1 - c_2) \approx
(c_9 + c_{10})/(c_1 + c_2)$~\cite{BBL}, neglecting a small exchange 
contribution~\cite{E} which vanishes at leading order in 
$1/m_b$ and $\alpha_s$~\cite{SCET}. SU(3) breaking effects on (\ref{T+C}), 
calculated by using generalized factorization~\cite{MN}, were found to be 
about 10$\%$ in the magnitude of ratio $(P_{EW} + P^C_{EW})/(T+C)$ and 
less than $5^\circ$ in its phase. Similar effects will be assumed in 
(\ref{C}), 
as estimated by similar considerations.  

The terms in the amplitudes (\ref{+-})-(\ref{0+}) multiplying $\lambda_t$ 
dominate the decay amplitudes because $|\lambda_u/\lambda_t|\approx 
0.02$. The penguin amplitude $P_{tc}$ is pure isosinglet, thus 
contributing equally to the two decay amplitudes involving a charged pion
and contributing a term smaller by factor $\s$ to the two amplitudes involving
$\pi^0$. Dominance by $P_{tc}$ is exhibited clearly by the four $K\pi$ 
branching ratios in Table I which obey Eq.~(\ref{Gamma}) to a reasonable 
approximation. (The effect of a lifetime difference between $B^+$ 
and $B^0$ will be discussed later.)
All other terms in (\ref{+-})-(\ref{0+}) are smaller 
than $\lambda_tP_{tc}$ and may be considered subdominant. 

Using Eq.~(\ref{T+C}) and noting that $T+C$ dominates the amplitude of 
$B^+\to\pi^+\pi^0$~\cite{GHLR}, the measured ratio of branching ratios 
$\b(\pi^+\pi^0)/\b(K^0\pi^+)$ shows that the higher order 
electroweak amplitude $P_{EW} + P^C_{EW}$  
is indeed much smaller than $P_{tc}$~\cite{NR,GLR},
\beq
\frac{|P_{EW} + P^C_{EW}|}{|P_{tc}|} \approx \frac{3}{2}\frac{|c_9 + c_{10}|}
{|c_1 + c_2|}\frac{\s f_K}{f_\pi}\frac{|V_{cb}|}{|V_{ud}V_{ub}|}
\sqrt{\frac{\b(\pi^+\pi^0)}{\b(K^0\pi^+)}} \approx 0.11~,
\eeq
where $f_\pi$ and $f_K$ are meson decay constants, and a value
$\b(\pi^+\pi^0) = (5.5\pm 0.6)\times 10^{-6}$~\cite{HFAG} was used. 

We will not assume color suppression for $C$ and $P^C_{EW}$,
nor will we assume that $P_{uc}$ is smaller 
than $T$ or $C$. That is, the triplet of amplitudes $(T, C, P_{uc})$ and 
the doublet $(P_{EW}, P^C_{EW})$ could each consist of amplitudes with 
comparable magnitudes. Several questions have been raised recently 
concerning these relative magnitudes~\cite{Yoshi,KpiSR,CGRS,BHLDS,BF} in 
view of an apparent disagreement with a hierarchy assumption~\cite{GHLR}
$|C| \sim 0.2|T|,~|P^C_{EW}| \sim 0.2|P_{EW}|$ and 
with calculations 
in QCD~\cite{BBNS,KLS}. We will make use of the fact that the amplitude 
$A$ and the strong phase of $C/T$ vanish to leading order in 
$1/m_b$ and $\alpha_s$~\cite{SCET}. We note that a small value of 
${\rm Arg}(C/T)$ is not favored by a global SU(3) fit to all $B$ meson 
decays into two charmless pseudoscalars~\cite{CGRS}, although the 
error on the output value of this phase is still very large. While the fit 
assumes common magnitudes and strong phases for SU(3) amplitudes in 
$B\to K\pi$ and $B\to \pi\pi$ decays, our assumption about SU(3) in 
Eqs.~(\ref{T+C})-(\ref{C}) is restricted to $B\to K\pi$. As mentioned,
these SU(3) breaking effects have been calculated to be very small
implying $|{\rm Arg}[(P_{EW}+P^C_{EW})/(T+C)]| < 5^\circ$.

Direct CP asymmetries in $B\to K\pi$ processes occur through the interference
of two terms in the amplitudes involving different CKM factors, $\lambda_t$
and $\lambda_u$, and different strong phases. Using the definition 
(\ref{Deltadef}) we find
\bea
\Delta(K^+\pi^-) & = & {\rm Im}\left [(P_{tc}+\frac{2}{3}P^C_{EW})
(T + P_{uc})^*\right ]I~,\\ 
2\Delta(K^+\pi^0) & = & {\rm Im}\left [(P_{tc}+ P_{EW} + 
\frac{2}{3}P^C_{EW})(T + C + A + P_{uc})^*\right ]I~,\\ 
2\Delta(K^0\pi^0) & = & {\rm Im}\left [(P_{tc} - P_{EW} - 
\frac{1}{3}P^C_{EW})(-C + P_{uc})^*\right ]I~,\\
\Delta(K^0\pi^+) & = & {\rm Im}\left [(P_{tc}- \frac{1}{3}P^C_{EW})
(A + P_{uc})^*\right ]I~,
\eea 
where $I=4\,{\rm Im}(\lambda_t\lambda^*_u)$ is a common CKM factor.

Combining the four CP rate differences by defining a 
difference $\delta_{K\pi}$ between pairs involving charged and neutral pions,
\beq
\delta_{K\pi}\equiv \Delta(K^+\pi^-) + \Delta(K^0\pi^+) - 
2\Delta(K^+\pi^0) - 2\Delta(K^0\pi^0)~,
\eeq 
we find
\beq\label{delta}
\delta_{K\pi} = -{\rm Im}\left [(P_{EW} + P^C_{EW})(T + C)^* + 
(P_{EW}C^* - P^C_{EW}T^*) + (P_{EW} + P^C_{EW})A^*\right ]I~.
\eeq
All terms involving the dominant $P_{tc}$ term cancel in $\delta_{K\pi}$.
This follows from isospin symmetry~\cite{AS}, as these terms describe
an interference of $P_{tc}$ with a term multiplying $\lambda_u$ in 
a combination which vanishes by Eqs.~(\ref{I-1})-(\ref{I-4}),
\beq\label{comb}
-A(K^+\pi^-) + A(K^0\pi^+) + \s A(K^+\pi^0) - \s A(K^0\pi^0) = 0~.
\eeq
 
All the terms on the right-hand-side of (\ref{delta}) involve EWP
amplitudes and are thus suppressed relative to corresponding terms 
involving $P_{tc}$ by 
about an order of magnitude. The first term vanishes in the 
SU(3) limit because it involves an interference of two contributions 
which carry a common strong phase by (\ref{T+C}). A potential SU(3) 
breaking strong phase difference, $|{\rm Arg}[(P_{EW}+P^C_{EW})/(T+C)]| < 
5^\circ$~\cite{MN}, suppresses this term by at least an order of 
magnitude. The second term
vanishes in the SU(3) limit at leading order in $1/m_b$ and $\alpha_s$,
as can be seen by using (\ref{T+C})-(\ref{C}) and ${\rm Arg}(C/T) 
\approx 0$. This implies a suppression of this term
either by an order of magnitude from SU(3) breaking,   
or by $1/m_b$ and $\alpha_s$.  
The last term on the right-hand-side involves an 
interference between two subdominant amplitudes, $P_{EW} + P^C_{EW}$ 
and $A$, each of which is suppressed relative to corresponding leading 
terms, $P_{tc}$ and $T$, respectively.
Since all three terms on the right-hand-side of (\ref{delta}) are doubly 
suppressed relative to $\Delta(K^+\pi^-)$ by two factors each about 
an order of magnitude, we expect the ratio
$\delta_{K\pi}/\Delta(K^+\pi^-)$ to be at most several percent. Therefore, 
one may safely take $\delta_{K\pi} =0$ which proves (\ref{DSR}).

The proposed sum rule may be written in terms of CP asymmetries, taking
into account differences among the four $B\to K\pi$ CP-averaged branching 
ratios and the $B^+$ to $B^0$ lifetime ratio $\tau_+/\tau_0 = 1.076 \pm 
0.008$~\cite{HFAG}:
\bea\label{exact}
A_{\rm CP}(K^+\pi^-) & + & A_{\rm CP}(K^0\pi^+)\frac{\b(K^0\pi^+)}
{\b(K^+\pi^-)}\frac{\tau_0}{\tau_+} \nonumber \\
& = & A_{\rm CP}(K^+\pi^0)\frac{2\b(K^+\pi^0)}{\b(K^+\pi^-)}\frac{\tau_0}
{\tau_+} + A_{\rm CP}(K^0\pi^0)\frac{2\b(K^0\pi^0)}{\b(K^+\pi^-)}~.
\eea
Using branching ratios from Table I, we predict a negative CP asymmetry 
in $B^0\to K^0\pi^0$ in terms of the other asymmetries,
\beq
A_{\rm CP}(K^0\pi^0) = -0.17 \pm 0.06~.
\eeq
This value is not inconsistent with the average measured value in Table I. 
Alternatively, the approximate sum rule (\ref{SR}) among CP asymmetries 
reads in terms of corresponding current measurements, 
\beq
(-0.120 \pm 0.019) + (-0.02 \pm 0.04) \approx (0.05 \pm 0.04) + 
(0.02 \pm 0.13)~.
\eeq
While central values on the two sides have opposite signs, errors in 
the asymmetries (in particular that in $B^0\to K^0\pi^0$) must be 
reduced before claiming a discrepancy.

The proposed sum rule (\ref{DSR})
makes no assumption about the smallness of the amplitudes $P_{uc}$ and $C$
relative to $T$, or about their strong phases relative to that of the 
dominant $P_{tc}$ amplitude. The contribution of $P_{uc}$ to the asymmetries 
has been neglected in a sum rule suggested recently when studying $b\to s$ 
penguin amplitudes in $B$ meson decays into two pseudoscalars~\cite{bspeng}. 
A $P_{uc}$ contribution comparable to $T$ would be observed by a nonzero 
$A_{\rm CP}(K^0\pi^+)$, unless the strong phase of $P_{uc}$ relative to 
$P_{tc}$ is very small. A sizable $P_{uc}$ comparable to $T$ is an output
of a global SU(3) fit to $B\to K\pi$ and $B\to\pi\pi$ 
decays~\cite{CGRS}. Bounds on $A_{\rm CP}(K^0\pi^+)$ derived from Table 
I favor a small relative phase between $P_{tc}$ and a sizable $P_{uc}$. 
Another output of the fit, a large amplitude $C$ comparable to $T$, also
obtained in separate analyses of $B\to K\pi$~\cite{BHLDS} and 
$B\to\pi\pi$~\cite{BF}, provides a simple interpretation for the failure 
of the oversimplified relation~(\ref{naive}) which had assumed $C$ to be 
color-suppressed. 
  
To conclude, we have shown that direct CP asymmetries in the four $B\to K\pi$
decay processes obey the sum rule (\ref{DSR}) within several percent, or the 
sum rule (\ref{SR}) in the approximation of equal rates in (\ref{Gamma}). 
Isospin and flavor SU(3) symmetries have been used to relate leading QCD 
penguin and subleading electroweak penguin terms in the sum rule, respectively.
While we assumed a suppression of an annihilation amplitude relative to a 
color-allowed tree amplitude and a suppression of ${\rm Arg}(C/T)$, no 
assumption was made about the magnitudes of 
color-suppressed tree, electroweak penguin amplitudes and a term $P_{uc}$ 
associated with intermediate $u$ and $c$ quarks. A violation of 
the sum rule would provide evidence for New Physics in 
$b\to s\bar qq$ transitions. The most likely interpretation of the origin of
a potential violation would be an anomalous $\Delta I=1$ operator in the 
effective Hamiltonian. A generalization of our argument 
using~(\ref{comb}) implies that
in the isospin symmetry limit contributions to CP asymmetries
from any $\Delta I=0$ operator cancel in the sum rule.
 
\bigskip
I wish to thank the SLAC theory group for its very kind hospitality.
I am grateful to Helen Quinn, Jonathan Rosner and Denis Suprun for 
very useful discussions. 
This work was supported in part by the Department of Energy contract 
DE-AC02-76SF00515, by the Israel Science Foundation founded 
by the Israel Academy of Science and Humanities, Grant No. 1052/04, and 
by the German-Israeli Foundation for Scientific Research and Development, 
Grant No. I-781-55.14/2003. 

\bigskip
\def \ajp#1#2#3{Am.\ J. Phys.\ {\bf#1}, #2 (#3)}
\def \apny#1#2#3{Ann.\ Phys.\ (N.Y.) {\bf#1}, #2 (#3)}
\def \app#1#2#3{Acta Phys.\ Polonica {\bf#1} (#3) #2}
\def \arnps#1#2#3{Ann.\ Rev.\ Nucl.\ Part.\ Sci.\ {\bf#1}, #2 (#3)}
\def \art{and references therein}
\def \cmts#1#2#3{Comments on Nucl.\ Part.\ Phys.\ {\bf#1}, #2 (#3)}
\def \cn{Collaboration}
\def \cp89{{\it CP Violation,} edited by C. Jarlskog (World Scientific,
Singapore, 1989)}
\def \econf#1#2#3{Electronic Conference Proceedings {\bf#1}, #2 (#3)}
\def \efi{Enrico Fermi Institute Report No.}
\def \epjc#1#2#3{Eur.\ Phys.\ J.\ C {\bf#1}, #2 (#3)}
\def \ib{{\it ibid.}~}
\def \ibj#1#2#3{~{\bf#1}, #2 (#3)}
\def \ijmpa#1#2#3{Int.\ J.\ Mod.\ Phys.\ A {\bf#1}, #2 (#3)}
\def \ite{{\it et al.}}
\def \jhep#1#2#3{JHEP {\bf#1} (#3) #2}
\def \jpb#1#2#3{J.\ Phys.\ B {\bf#1}, #2 (#3)}
\def \mpla#1#2#3{Mod.\ Phys.\ Lett.\ A {\bf#1} (#3) #2}
\def \nat#1#2#3{Nature {\bf#1}, #2 (#3)}
\def \nc#1#2#3{Nuovo Cim.\ {\bf#1}, #2 (#3)}
\def \nima#1#2#3{Nucl.\ Instr.\ Meth.\ A {\bf#1}, #2 (#3)}
\def \npb#1#2#3{Nucl.\ Phys.\ B~{\bf#1} (#3) #2}
\def \npps#1#2#3{Nucl.\ Phys.\ Proc.\ Suppl.\ {\bf#1}, #2 (#3)}
\def \PDG{Particle Data Group, K. Hagiwara \ite, 
\prd{66}{010001}{2002}}
\def \pisma#1#2#3#4{Pis'ma Zh.\ Eksp.\ Teor.\ Fiz.\ {\bf#1}, 
#2 (#3) [JETP
Lett.\ {\bf#1}, #4 (#3)]}
\def \pl#1#2#3{Phys.\ Lett.\ {\bf#1} (#3) #2}
\def \pla#1#2#3{Phys.\ Lett.\ A {\bf#1}, #2 (#3)}
\def \plb#1#2#3{Phys.\ Lett.\ B {\bf#1} (#3) #2}
\def \prl#1#2#3{Phys.\ Rev.\ Lett.\ {\bf#1} (#3)  #2}
\def \prd#1#2#3{Phys.\ Rev.\ D\ {\bf#1} (#3) #2}
\def \prp#1#2#3{Phys.\ Rep.\ {\bf#1}, #2 (#3)}
\def \ptp#1#2#3{Prog.\ Theor.\ Phys.\ {\bf#1}, #2 (#3)}
\def \rmp#1#2#3{Rev.\ Mod.\ Phys.\ {\bf#1} (#3) #2}
\def \yaf#1#2#3#4{Yad.\ Fiz.\ {\bf#1}, #2 (#3) [Sov.\ 
J.\ Nucl.\ Phys.\
{\bf #1}, #4 (#3)]}
\def \zhetf#1#2#3#4#5#6{Zh.\ Eksp.\ Teor.\ Fiz.\ {\bf #1}, 
#2 (#3) [Sov.\
Phys.\ - JETP {\bf #4}, #5 (#6)]}
\def \zpc#1#2#3{Zeit.\ Phys.\ C {\bf#1}, #2 (#3)}
\def \zpd#1#2#3{Zeit.\ Phys.\ D {\bf#1}, #2 (#3)}


\begin{thebibliography}{99}
\bibitem{E} Evidence for suppression by an order of magnitude of an 
exchange amplitude relative to a tree amplitude exists in $B^0\to D^-_s 
K^+$~\cite{PDG}. 
The measured suppression factor in the charmless decay process 
$B^0 \to K^+ K^-$, for which a strict upper limit has been recently
obtained~\cite{Ford,K+K-}, has not yet reached this level.
\end{thebibliography}
\end{document}